\documentclass[letterpaper,twocolumn,showpacs,preprintnumbers,amsmath,amssymb,floatfix,superscriptaddress,pre]{revtex4}

\usepackage{psfrag,color}
\usepackage{graphicx}  
\usepackage{dcolumn}   
\usepackage{bm}        
\usepackage{epsfig}
\usepackage{textcomp}
\usepackage[sort&compress]{natbib}

\newcommand{\secn}[1]{\section{#1}}
\newcommand{\subsecn}[1]{\subsection{#1}}
\newcommand{\subsecnnn}[1]{\subsection*{#1}}

\begin{document}

\title{
High temperature correlation functions: universality, 
extraction of exchange interactions, divergent correlation lengths
and generalized Debye length scales}

\author{Saurish Chakrabarty}
\affiliation{Department of Physics and Center for Materials Innovation, Washington University in St Louis, MO 63130.}

\author{Zohar Nussinov}
\affiliation{Department of Physics and Center for Materials Innovation, Washington University in St Louis, MO 63130.}
\affiliation{Kavli Institute for Theoretical Physics, Santa Barbara, CA 93106}

\date{\today}

\begin{abstract}
We derive a universal
form for the correlation function of
general $n$ component systems in the limit
of high temperatures or weak coupling.
This enables the extraction of effective microscopic interactions
from measured high temperature correlation functions. 
We find that in systems with long range interactions,
there exist diverging correlation lengths
with amplitudes that tend to zero
in the high temperature limit. 
For general systems with disparate long range interactions,
we introduce the notion of generalized Debye length (and time) scales and 
further relate it to the divergence of the largest correlation
length in the high temperature (or weak coupling) limit. 
\end{abstract}

\pacs{05.50.+q, 75.10.-b, 75.10.Hk}
\maketitle
\secn{Introduction}
The study of  correlation functions in systems with multi-component fields is an extremely
general problem having incarnations that range from systems in condensed matter
physics to fundamental field theories.
One of the cornerstones  of field theories and the study of 
critical phenomena is the recognition of the universality that underlies general
systems.
This enables a unified understanding
and potent tools of analysis \cite{ma_book, wilson, fisher}.
Any system generally displays a disordered high temperature $(T)$ fixed point.
Most of the research to date focused on 
the behavior of myriad systems at and in
the vicinity of various finite temperature transitions.
In this work, we will focus on high temperature behavior and
illustrate that a simple form
of two-point correlation functions is universally exact 
for rather general systems. 
This will enable us to make several striking observations.
In particular, we will demonstrate
that in contrast to common intuition, general
systems with long range interactions have a 
correlation length that increases monotonically
with temperature as $T\to\infty$. As they must, 
however,
the correlations
decay monotonically with temperature  (as the corresponding
amplitudes decay algebraically with temperature).
There have been no earlier reports of diverging correlation lengths
at high temperature. 
A thermally increasing length-scale 
of a seemingly very different sort appears in plasmas \cite{debye_huckel}.
The Debye length, the distance over 
which screening occurs in a plasma,
diverges, at high temperature, as $\lambda_D \propto \sqrt{T}$.
We introduce the notion of a {\em generalized Debye length} associated
with disparate long range interactions (including confining interactions) and show that such screening lengths are rather
general. 

Many early works investigated the high temperature disordered phase via a high temperature series expansion  \cite{stanleykaplan}
with an eye towards systems with
short  range interactions. In this paper, we report on our universal
result for the Fourier transformed correlation function 
for systems with general pair interactions. 
As it must, for nearest neighbor interactions, our correlation 
function agrees with what is suggested by standard approximate methods
(e.g., the Ornstein-Zernike (OZ) correlation function that may be derived by many approximate schemes
\cite{huang}). Our work
places such approximate results on a more rigorous footing and, perhaps most notably,
enables us to go far beyond standard
short range interactions to find rather surprising results. Our derivations will be done for spin and other general
lattice systems with multi-component fields. However, as illustrated later, our results
also pertain to continuum theories. 

\secn{Outline}
In Section \ref{sos}, we introduce the systems we study 
(general multi-component spin systems on a lattice; later sections generalize our result to 
other arenas -- fluids, Bose/Fermi systems and so on).
Section \ref{corrfunc} contains a derivation of our main result about 
the universal form of the correlation function in the high temperature limit.
In Section \ref{htcl}, we comment on how the correlation lengths in a system behave in the high temperature limit.
Section \ref{gdl} introduces a generalized Debye length.
In Section \ref{gen}, we present some generalizations of our result to systems which are not covered in Section \ref{sos}.
Section \ref{approx} outlines standard approximate techniques used to obtain our result.
We give our concluding remarks in  Section \ref{conc}.

In Appendix \ref{htseries}, we show how to obtain a full high temperature series expansion of the correlation function 
to arbitrary order. In Appendix \ref{debye_rel}, we relate the generalized Debye length over which long range 
interactions are screened to the diverging correlation length present in the high temperature limit.

\secn{Systems of study}\label{sos}
We consider a translationally invariant system with the Hamiltonian 
\begin{eqnarray}
H = \frac{1}{2} \sum_{\vec{x} \neq \vec{y}}V(|\vec{x}-\vec{y}|)\vec{S}(\vec{x})\cdot\vec{S}(\vec{y}). 
\label{Ham}
\end{eqnarray} 
The sites $\vec{x}$ and $\vec{y}$ lie on a $d$-dimensional hyper-cubic lattice
with $N$ sites having unit lattice constant.
The quantities $\{\vec{S}(\vec{x})\}$ portray $n$-component spins (``$O(n)$ spins'') 
or general fields where $|\vec{S}(\vec{x})|^2=n$
at all lattice sites $\vec{x}$. The normalization is adopted from \cite{stanley}. 
The case of $n=1$ corresponds to
Ising spins,  $n=2$ to XY spins, 
and so on.
We assume that whenever the interaction kernel $V(x)$ has a long range component,
that component (unless stated otherwise) will always have
some {\em finite screening}, however small.
This restriction is imposed to avoid well-known 
difficulties in taking thermodynamic limits
in long range systems. 
In what follows, $v(\vec{k})$
and $s_i(\vec{k})$ are the Fourier transforms of
$V(|\vec{x}-\vec{y}|)$ and $S_i(\vec{x})$.
With this, Eq.(\ref{Ham}) reads
$
H=\frac{1}{2N}\sum_{\vec{k}}v(\vec{k})\vec{s}(\vec{k})\cdot\vec{s}(-\vec{k}),
$
up to an innocuous constant.
Throughout, we employ
the Fourier transform convention of 
$a(\vec{k})=\sum_{\vec{x}}A(\vec{x})e^{i\vec{k}\cdot\vec{x}}$ (and
$A(\vec{x})=\frac{1}{N}\sum_{\vec{k}}a(\vec{k})e^{-i\vec{k}\cdot\vec{x}}$).

\secn{The universal form of the high temperature correlation functions}\label{corrfunc}
 We now derive a universal form for the correlation function at high temperature.
 As in any other calculation with Boltzmann weights, the high temperature limit 
 is synonymous with weak coupling.  
 Initially, we follow standard procedures 
 and examine a continuous but exact dual theory. High $T$ (or weak coupling) in the original 
 theory corresponds to strong coupling in the dual
 theory. We will then proceed to examine
 the consequences of the dual theory at high temperature where the strong coupling
 interaction term dominates over other non-universal terms that depend, e.g., on
 the number of components in the original theory. This enables an 
 analysis with general results. Unlike most treatments that focus on
the character of various phases
and intervening transitions, 
our interest here is strictly in the 
{\em high temperature limit of the correlation
functions} in rather general theories of Eq.(\ref{Ham}). Our aims 
are {\bf (i)} to make conclusions concerning systems with
long range interactions rigorous and {\bf (ii)} to extract microscopic
interactions from measurements.
It is notable that due to convergence time constraints 
many numerical approaches, e.g., \cite{kob}, compare candidate potentials with
experimental data at high temperature (above the melting
temperatures)
where the approach that we will outline is best suited. We will 
perform a transformation to a continuous but
exact dual theory where the high temperature character
of the original theory can be directly examined.

 We augment
the right hand side of  Eq.(\ref{Ham}) by  $ [-\sum_{\vec{x}}\vec{h}(\vec{x})\cdot\vec{S}(\vec{x})]$ 
and differentiate in the limit $\vec{h} \to 0$ to obtain correlation functions in the usual way.
\begin{eqnarray}
G(\vec{x}-\vec{y})
&=&  \frac{1}{n}\left\langle\vec{S}(\vec{x})\cdot\vec{S}(\vec{y})\right\rangle\nonumber\\
&=& \lim_{h \to 0} \frac{1}{n\beta^2Z}\sum_{i=1}^n \frac{\delta^2 Z}{\delta  h_i(\vec{x}) \delta h_i(\vec{y})},
\label{ghx}
\end{eqnarray}
with $Z$ the partition function in the presence of the external field $\vec{h}$.
By spin normalization,
$
G(\vec{x})=1 \mbox{ for } \vec{x}=0.
$
The index $i=1,2,..., n$ labels the $n$ internal spin (or field) components. 
The partition function
$
Z=\mbox{Tr}_S \left[\exp\left(-\frac{\beta}{2N}\sum_{\vec{k}}v(\vec{k})|\vec{s}(\vec{k})|^2+\beta\sum_{\vec{x}}\vec{h}(\vec{x})\cdot\vec{S}(\vec{x})\right)\right].
$
The subscript $S$ denotes the trace with respect to the spins.
Using the Hubbard-Stratonovich (HS) transformation, \cite{SRL,ivanchenko}
we introduce the dual variables $\{\vec{\eta}(\vec{x})\}$
and rewrite the partition function as
\begin{eqnarray}
Z=\mbox{Tr}_S \left[\prod_{\vec{k},i}\left([2\pi(-v(\vec{k}))]^{-1/2}\right.\right.\times\ \ \ \ \ \ \ \ \ \ \ \ \ \ \ \ \ \ \ \ \ \nonumber\\
\left.\int_{-\infty}^{\infty}d\eta_i(\vec{k})e^{\frac{N}{2\beta
        v(\vec{k})}\left|\eta_i(\vec{k})\right|^2+\eta_i(\vec{k})s_i(-\vec{k})}\right)
\prod_{\vec{x}}\left.e^{\beta\vec{h}(\vec{x})\cdot\vec{S}(\vec{x})}\right]\label{part1}\\
={\cal{N}}~ \mbox{Tr}_S\left[\int
  d^{Nn}\eta\exp\left(\frac{N^2}{2\beta}\sum_{\vec{x},\vec{y}}V^{-1}(\vec{x}-\vec{y})\vec{\eta}(\vec{x})\cdot\vec{\eta}(\vec{y})\right.\right.\nonumber\\
+\left.\left.N\sum_{\vec{x}}\vec{\eta}(\vec{x})\cdot\vec{S}(\vec{x})+\beta\sum_{\vec{x}}\vec{h}(\vec{x})\cdot\vec{S}(\vec{x})\right)\right]\label{treq},
\end{eqnarray}
with $V^{-1}(\vec{x})$ the inverse Fourier transform of  
$1/v(\vec{k})$ and ${\cal N}$ a numerical prefactor.
The physical motivation in performing the duality to the HS variables is that
we wish to retain the exact character of the theory (i.e., the exact form of the interactions and 
the $O(n)$ constraints concerning 
the spin normalization at all lattice sites). It is for this reason that we do not resort to a continuum approximation 
(such as that of the canonical $\phi^{4}$ theory that we will discuss for comparison later on)
where normalization is not present.
Another reason to choose to work in the dual space is the correspondence with 
field 
theories 
that, in the dual space,
becomes clearer in the high temperature limit (in which the quartic term of the $\phi^4$ theories becomes irrelevant).
Further details 
are in \cite{explain_shift}.
For $O(n)$ spins, 
\begin{eqnarray}
Z&=&{\cal{N}}'\int
d^{Nn}\eta\nonumber\\
& &\left[\exp\left(\frac{N^2}{2\beta}\sum_{\vec{x},\vec{y}}V^{-1}(\vec{x}-\vec{y})\vec{\eta}(\vec{x})\cdot\vec{\eta}(\vec{y})\right.\right)\times\nonumber\\
& &\left.\prod_{\vec{x}}  \frac{I_{n/2-1}(\sqrt{n}|N\vec{\eta}(\vec{x})+\beta\vec{h}(\vec{x})|)}{(\sqrt{n}|N\vec{\eta}(\vec{x})+\beta\vec{h}(\vec{x})|)^{n/2-1}}  \right].
\label{Zeta}
\end{eqnarray}
The second factor in Eq.(\ref{Zeta}) originates from the trace over the spins
and as such embodies the $O(n)$ constraints (the trace  in Eq.(\ref{treq}) is performed
over all configurations with $(\vec{S}(\vec{x}))^{2}=n$ at all sites $\vec{x}$).  
Here, $I_{\nu}(x)$ is the modified Bessel function of the first kind. 
In the Ising ($n=1$) case, the argument of the product in Eq.(\ref{Zeta}) is a hyperbolic cosine.
Up to an innocuous additive constant, Eq.(\ref{Zeta}) corresponds to 
the dual Hamiltonian,
\begin{eqnarray}
H_{d}&=&-\frac{N^2}{2\beta^2}\sum_{\vec{x},\vec{y}}V^{-1}(\vec{x}-\vec{y})\vec{\eta}(\vec{x})\cdot\vec{\eta}(\vec{y})\nonumber\\
& &-\frac{1}{\beta}\sum_{\vec{x}}\ln\left(\frac{I_{n/2-1}(\sqrt{n}|N\vec{\eta}(\vec{x})+\beta\vec{h}(\vec{x})|)}{(\sqrt{n}|N\vec{\eta}(\vec{x})+\beta\vec{h}(\vec{x})|)^{n/2-1}}\right).
\label{dual}
\end{eqnarray}
Our interest is in the $h \to 0$ limit. The first term in Eq. (\ref{dual}) is the same 
for all $n$. This term dominates, at low $\beta$, over the (second) $n$ dependent term.
As we will see, this dominance will enable us to get universal results for all $n$.
From Eq.(\ref {ghx}), and the identity
\begin{eqnarray}
\frac{d}{dx}\left[\frac{I_{\nu}(x)}{x^{\nu}}\right]=\frac{I_{\nu+1}(x)}{x^{\nu}},\nonumber
\end{eqnarray}
we find that
\begin{eqnarray}
&&G(\vec{x}-\vec{y})
=\delta_{\vec{x},\vec{y}}+
(1-\delta_{\vec{x},\vec{y}})\left\langle
\frac{\vec{\eta}(\vec{x})\cdot\vec{\eta}(\vec{y})}
{|\vec{\eta}(\vec{x})||\vec{\eta}(\vec{y})|}\right.\times\nonumber\\
& &\left.
\frac{I_{n/2}(N\sqrt{n}|\vec{\eta}(\vec{x})|)I_{n/2}(N\sqrt{n}|\vec{\eta}(\vec{y})|)}
{I_{n/2-1}(N\sqrt{n}|\vec{\eta}(\vec{x})|)I_{n/2-1}(N\sqrt{n}|\vec{\eta}(\vec{y})|)}
\right\rangle_{d},
\label{gtan}
\end{eqnarray}
where the average ($\langle . \rangle_{d}$) performed with the weights $\exp(-\beta H_{d})$. 
Now, here is a crucial idea regarding our exact dual forms. 
From Eq.(\ref{part1}), {\em at high temperature}, the variables
$\eta_i(\vec{k})$ strictly have sharply peaked Gaussian distributions of
variance,
\begin{eqnarray}
\left\langle\left|\eta_i(\vec{k})\right|^2\right\rangle_{d} \approx\frac{-\beta
  v(\vec{k})}{N} \mbox{ as }\beta\to0.
  \label{gv}
\end{eqnarray}
Importantly, this variance tends to zero as $\beta\to0$.
By Parseval's theorem and translational invariance, 
\begin{eqnarray}
\left\langle \left(\eta_i(\vec{x})\right)^2\right\rangle_{d} 
&&= \frac{1}{N}\sum_{\vec{x}}\left\langle \left(\eta_i(\vec{x})\right)^2\right\rangle_{d} \nonumber\\
&&= \frac{1}{N^2}\sum_{\vec{k}}\left\langle \left|\eta_i(\vec{k})\right|^2\right\rangle_{d} 
\approx-\beta V(0)/N^2\nonumber
\end{eqnarray}
Thus, at high temperature,
$\langle (\eta_i (\vec{x}))^2 \rangle \ll1$.
It is therefore useful to perform a series expansion in the dual variables $\eta$ and this would give rise 
to a high temperature series expansion in the correlation function.
\begin{eqnarray}
H_{d}&=&-\frac{N^2}{2\beta^2}\sum_{\vec{x},\vec{y}}V^{-1}(\vec{x}-\vec{y})\vec{\eta}(\vec{x})\cdot\vec{\eta}(\vec{y})-\frac{N^2}{2\beta}\sum_{\vec{x},i}\eta_i(\vec{x})^2, \nonumber\\
&=&-\frac{N}{2\beta^2}\sum_{\vec{k},i}\frac{1}{v(\vec{k})}\left|\eta_i(\vec{k})\right|^2-\frac{N}{2\beta}\sum_{\vec{k},i}\left|\eta_i(\vec{k})\right|^2,
\label{HG}
\end{eqnarray}
with errors of ${\cal O}(1/T)$.
Expanding Eq.(\ref{gtan}) to  ${\cal O}(1/T^2)$, 
\begin{eqnarray}
\boxed{
G(\vec{k})=\frac{k_BT}{v(\vec{k})+k_BT}+\frac{1}{N}\sum_{\vec{k'}}\frac{v(\vec{k'})}{v(\vec{k'})+k_BT}.
}
\label{gkon'}
\end{eqnarray}
Eq.(\ref{gkon'}) leads to counter-intuitive consequences for
systems with long-range interactions.
The second term in Eq.(\ref{gkon'}) is independent of $\vec{k}$
and ensures that $G(\vec{x})=1$ for $\vec{x}=0$.
Inverting this result enables us to find the microscopic (spin exchange or other) interactions
from the knowledge of the high temperature correlation function.
We thus flesh out (and further generalize for multicomponent systems such as spins) 
the mathematical uniqueness theorem of Henderson for fluids \cite{henderson}
for which a known correlation function $G(\vec{x})$ leads to a known pair potential function $V(\vec{x})$
up to an innocuous constant.
Eq.(\ref{gkon'}) leads to a correlation function which is independent of $V(0)$. Therefore,
we can shift $v(\vec{k})$ for all $\vec{k}$'s by an arbitrary constant or equivalently set $V(0)$
to an arbitrary constant.
To ${\cal{O}}(1/T)$, for $V(0)=0$, we have, 
\begin{eqnarray}
v(\vec{k})=\frac{k_BT}{G(\vec{k})}-\frac{1}{N}\sum_{\vec{k'}}\frac{k_BT}{G(\vec{k'})}.
\label{vk}
\end{eqnarray}
The leading term of this expression for $v(\vec{k})$ does not scale with $T$.
This is because $(1-G(\vec{k}))\propto1/T$ at high temperature.
Correlation functions obtained from experimental data can be plugged into the right hand side to obtain the 
effective pair potentials. 
Alternatively, in real space,
for $\vec{x}\neq0$,
\begin{eqnarray}
\boxed{
V(\vec{x})=-k_BTG(\vec{x})+k_BT\sum_{\vec{x'}\neq0,\vec{x}}G(\vec{x'})G(\vec{x}-\vec{x'})
}
\label{Vx}
\end{eqnarray}
Note that the two terms in Eq.(\ref{Vx}) are ${\cal{O}}(1)$ and
${\cal{O}}(1/T)$ respectively, since $G(\vec{x})$  is proportional to $1/T$ at high temperature for $\vec{x}\neq0$.
Extension to higher orders may enable better comparison to experimental or numerical data.
Our expansion is analytic in the high temperature phase
(i.e., so long as no transitions are encountered as $1/T$ is increased from zero). 
The Gaussian form of Eq.(\ref{HG}) similarly leads to
the free energy density,
\begin{eqnarray}
\label{free}
F = \frac{k_{B}T}{2N} \sum_{\vec{k}} \ln\left|\frac{k_BT}{v(\vec{k})} +1\right|+{\cal{O}}(1/T).
\end{eqnarray}
Armed with Eq.(\ref{HG}, \ref{gkon'}), we can compute any correlation
function with the aid of Wick's theorem. For example, for unequal $\vec{k_i}$s, we have,
$
\langle (\vec{s}(k_{1}) \cdot \vec{s}(-k_{1})) ... (\vec{s}(k_{m}) \cdot \vec{s}(-k_{m})) \rangle
 = (Nn)^m\prod_{i=1}^{m}G(\vec{k_i}).
$


It is straightforward to carry out a full high temperature series expansion of the correlation function to arbitrary order.
This is outlined in Appendix \ref{htseries}.
For example, to ${\cal O}(1/T^3)$, for $V(\vec{x}=0)=0$, 
the correlation function in real space, is given for $\vec{x}\neq0$ by,
\begin{eqnarray}
\label{g3v00}
G(\vec{x})&=&-\frac{V(\vec{x})}{k_BT}+\frac{1}{(k_BT)^2}\sum_{\vec{z}}V(\vec{z})V(\vec{x}-\vec{z})\nonumber\\
&&-\frac{1}{(k_BT)^3}\left[\sum_{\vec{y},\vec{z}}V(\vec{y})V(\vec{z})V(\vec{x}-\vec{y}-\vec{z})\right.\nonumber\\
&&-\left.2V(\vec{x})\sum_{\vec{z}}V(\vec{z})V(-\vec{z})+2\frac{(V(\vec{x}))^3}{n+2}\right].
\end{eqnarray}

\secn{High temperature correlation lengths} \label{htcl}
We now illustrate that  {\bf (i)}  in systems with short (or finite
range) interactions, the correlation length tends to
zero in the high temperature limit and {\bf(ii)} in systems 
with long range interactions \cite{what_is_long?}
the high temperature correlation
length tends to the screening length 
and diverges in the absence of
screening.
\subsecn{Decaying lengthscales}
We consider first the standard case of short range interactions. 
On a hyper-cubic lattice in $d$ spatial dimensions, nearest neighbor interactions have the lattice Laplacian
$\Delta(\vec{k}) = 2 \sum_{l=1}^{d} (1-\cos k_{l})$ ,  with $k_{l}$ the $l$-th 
Cartesian component of the wave-vector $\vec{k}$
as their Fourier transform. In the continuum (small $k$) limit,
$\Delta\sim |\vec{k}|^{2}$. Generally, in the continuum, arbitrary finite
range interactions of spatial range $p$ have $v(\vec{k}) \sim |\vec{k}|^{2p}$
with $p>0$ (and superposition of such terms thereof)
as their Fourier transform.  In general finite range
interactions, similar multi-nomials in
$(1-\cos k_{l})$
and in $k_{l}^{2}$ appear on the lattice
and the continuum respectively.  For simplicity, we consider
$v(\vec{k}) \sim |\vec{k}|^{2p}$.
Correlation lengths are determined
by the reciprocal of the imaginary part of poles
of Eq.(\ref{gkon'}),  $|$Im $\{k_* \}|^{-1}$. We then have
that in the complex $k$ plane, $(k_*)^{2p} \sim -{k_BT}$.
Poles are given by $k_*  \sim (k_{B} T)^{1/(2p)} \exp[ (2m+1)\pi i/(2p)]$ with $m=0,1, ..., 2p-1$.
Correlation lengths then tend to zero in the high temperature
limit as $\xi \sim T^{-1/(2p)}/|\sin (2m+1) \pi/(2p)|$ -- there are $p$ such correlation
lengths. Similarly, there are $p$ periodic modulation lengths
scaling as $L_{D} \sim 2 \pi  T^{-1/(2p)}/|\cos (2m+1) \pi/(2p)|$. The usual case of $p=1$ corresponds to
an infinite $L_{D}$ (i.e., spatially uniform (non-periodic) correlations) and $\xi \sim T^{-1/2}$. 
\subsecn{Diverging lengthscales}
The novelty arises in the high temperature limit of systems with long range interactions
where $v(\vec{k})$ diverges in
the small $k$ limit. Such a divergence enables the correlator
of Eq.(\ref{gkon'}) to have a pole at low $k$ and consequently, on Fourier transforming
to real space, to have a divergent correlation length. In the presence, 
of screening, $v(\vec{k})$ diverges and $G(k)$ has a pole
when the imaginary part of $k$ is equal to
the reciprocal of the screening length. The correlation length
then tends to the screening length at high temperature.
For concreteness, we consider generic screened interactions
where the Fourier transformed interaction kernel
$v_L(k) \sim \frac{1}{(k^{2}+\lambda^{-2})^{p'}}$
with $p'>0$ and $\lambda$ the screening length. 
Perusing the poles of Eq.(\ref{gkon'}),
we find that for all $p'$, the correlation lengths tend to the screening
length in the high temperature limit,
\begin{eqnarray}
\boxed{
\lim_{T \to \infty} \xi(T) = \lambda.
}
\label{xilam}
\end{eqnarray}
From Eq.(\ref{xilam}), when $\lambda$ becomes arbitrarily large, the {\em correlation length diverges}.
Physically, such correlations enable global ``charge neutrality''  \cite{neutral} for the corresponding 
long range interactions (Coulomb or other).
This general divergence of high temperature correlation lengths in systems with long range interactions
is related to the effective range of the interactions.
At high temperature, the correlation function matches the ``direct'' contribution,
$e^{-\beta V_{eff}(\vec{r})}-1\sim -\beta V_{eff}(\vec{r})$.
If the effective interactions between two fields have a range $\lambda$,
then that is reflected in the correlation length.
In Coulomb systems, the Debye length, $\lambda_D$ sets the range of the interactions
(for large distances, the interactions are screened).
As stated earlier, at high temperature, $\lambda_D$ diverges.
As seen by Fourier transforming Eq.(\ref{gkon'}),
though the imaginary part of the poles tends to zero
(and thus the correlation lengths diverge), the prefactor multiplying
$e^{-|\vec{x}|/\xi}$ is a monotonically
decaying function of $T$. Thus in the high temperature limit the real
space correlator $G(\vec{x})$ monotonically decays with temperature (as it must). 
For instance, for $p'=1$ in $d=3$ dimensions, the pair correlator
$G(x) \sim e^{-x/\lambda}/(Tx)$
tends, for any non-zero $x$, to zero as $T \to \infty$. \cite{saurish}
That is, {\em the amplitude vanishes} in the high temperature limit as
$(1/T)$. 
We find similar results
when we have more than
one interaction. For instance,
in the presence of both a short and a long range interaction,
(at least) two correlation lengths are found.
One correlation length (or, generally, set of correlation lengths) tends to zero
in the high temperature limit (as 
for systems with short range interactions)
while the other correlation length (or such set) tends
to the screening length (as we find 
for systems with long range interactions).
An example of a system where this can be observed is    
the screened ``Coulomb Frustrated Ferromagnet'', \cite{us, saurish, zohar}
given by the Hamiltonian
$H=[-J\sum_{ \langle \vec{x},\vec{y} \rangle}S(\vec{x})S(\vec{y})+Q\sum_{\vec{x} \neq \vec{y}}V_L(|\vec{x} - \vec{y}|)S(\vec{x})S(\vec{y})]$,
with  $J,Q>0$ and the long range interaction $V_L(x)=\frac{e^{-x/\lambda}}{x}$ in $d=3$ dimensions and
$V_L(x)=K_0(x/\lambda)$ in $d=2$
with $\lambda$ the screening length
and $K_{0}$ a modified Bessel function of the second kind.
Similar dipolar systems \cite{dipole1, dipole2, dipole3}
were considered. Apart from the usual correlation length that vanishes
in the high temperature limit,  we find an additional correlation
length that tends to the screening length $\lambda$.

\secn{Generalized Debye length (and time) scales}\label{gdl}
We now introduce the notion of generalized Debye length (and time) scales
that are applicable to general systems with effective or exact long range interactions. 
These extend the notion of a Debye length from Coulomb type system where
it is was first found.
If the Fourier space interaction kernel $v(k)$ 
in a system with long range interactions
is such that 
$\frac{1}{v(k)}$
is analytic at $k=0$, then the
system has a diverging correlation length, $\xi_{long}$ at high temperature.
To get the characteristic diverging lengthscales, we consider the self-consistent small $k$ solutions
to $k_BT/v(k)=-1$ for high temperature (which gives the poles in the correlation function).
Thus, as $T\to\infty$,
$\xi_{long}$ diverges as $\sqrt[p]{k_BT}$, where
$p$ is the order of the first non-zero term in the Taylor series expansion of 
$\frac{1}{v(k)}$
around $k=0$.
This divergent length-scale could be called the generalized Debye length.
If the long-range interactions in the system are of Coulomb type, then this corresponds to the usual
Debye length $\lambda_D$ where $p=2$.
A more common way to obtain this result is as follows.
Suppose we have our translationally invariant system which interacts via pairwise couplings
as in Eq.(\ref{Ham}). 
We can define a potential function for this system as,
\begin{eqnarray}
\phi(\vec{x})=\sum_{\vec{y},\vec{y}\neq\vec{x}}V(|\vec{x}-\vec{y}|)S(\vec{y}).
\end{eqnarray}
The ``charge'' $S(\vec{x})$ in the system is perturbed by an amount $\hat{S}(\vec{x})$ and we observe the
response $\hat{\phi}(\vec{x})$ in the potential function $\phi(\vec{x})$ assuming that we stay within the
regime of linear response.
We assume $S(\vec{x})$ follows a Boltzmann distribution, i.e., $S(\vec{x})=A\exp\left(-\beta C\phi(\vec{x})\right)$, where
$C$ is a constant depending on the system.
It follows that $\hat{S}(\vec{x})=-\beta C S(\vec{x}) \hat{\phi}(\vec{x})$.
At this point, we can ignore the fluctuations in $S(\vec{x})$ as it does not contribute to the leading order
term. Thus, $\hat{S}(\vec{x})=-\beta C S_0 \hat{\phi}(\vec{x})$, where $S_0=\langle S(\vec{x})\rangle$.
In Fourier space, this leads to the relation,
\begin{eqnarray}
\hat{\phi}(\vec{k})=-\beta C S_0 v(\vec{k})\hat{\phi}(\vec{k})
\end{eqnarray}
The modes with non-zero response are therefore given by, 
\begin{eqnarray}
\label{gd}
\boxed{-v(\vec{k})\propto k_BT.}
\end{eqnarray}
For a Coulomb system, these modes are given by $(-k^{-2})\propto k_BT$, 
yielding a correlation length $\lambda_D\propto\sqrt{k_BT}$.

As a brief aside, we remark that, repeating all of the above considerations
(and also those to be detailed anew in Section \ref{gl}), if an imaginary time action 
for a complex field $\psi$ has the 
form 
\begin{eqnarray}
\label{action}
S_{action} = \frac{1}{2} \int d\tau d\tau' d^{d}x d^{d}x' ~ \Big[  \psi (x, \tau)  \nonumber
\\ K(x-x', \tau-\tau') \psi(x',\tau') 
\Big]+ ...,
\end{eqnarray}
with the imaginary time coordinates $0 \le \tau, \tau' \le \beta$
with a kernel $K$ that is long range in space or imaginary time 
and the ellipsis denoting higher order terms (e.g., generic $|\psi|^4$ type terms)
or imposing additional constraints on the fields $\psi$ (such as normalization
that we have applied thus far for $O(n)$ systems)
then the associated Debye length (or imaginary time) scale
may diverge in the weak coupling (i.e., $K \to a K$ with $a \to 0^{+}$) limit. 
In analogous way, repeating all of the earlier calculations done thus
far for spatial correlations, we find that divergent correlation
times in the low coupling limit for systems with a kernel $K$
that is long range in $|\tau-\tau'|$.
An action such as that of Eq.(\ref{action}) may also describe a system
at the zero temperature limit (whence $\beta \to \infty$)
and the (imaginary) time scale is unbounded.

In Appendix \ref{debye_rel}, we will relate the divergence of
the {\em generalized Debye type length scales}
in the high temperature limit to a similar
divergence in the largest correlation
length in systems with long range
interactions.

\subsecnnn{Confining potentials}
We discussed long range interactions (with, in general, 
a screening which may be set to be arbitrarily small)
such as those that arise in plasma, dipolar systems, and other systems in condensed matter physics. 
In all of these 
systems, the long range potentials dropped monotonically with increasing distance. Formally, we may consider
generalizations which further encompass confining potentials such as those that capture the effective confining potentials 
in between quarks in quantum chromodynamics (QCD) as well as those between charges in one dimensional 
Coulomb systems (where the effective potentials associated with the electric flux tubes in one dimension
lead to linear potentials). The derivations that we carried throughout also hold in such cases. For instance,
in a one-dimensional Coulomb system, the associated linear potential $V(x) \sim |x|$ leads to the usual Coulomb Fourier
space kernel $v(k) \sim k^{-2}$. In general, for a potential $V(x) \sim |\vec{x}|^{-a}$ in $d$ spatial dimensions, 
the corresponding Fourier space
kernel is, as in the earlier case, 
$v(k) \sim |\vec{k}|^{-p}$, where
$p=d-a$. Following the earlier discussion, 
this leads, at asymptotically high temperatures (and for infinitesimal screening),
to correlation lengths that scale as $\xi \sim \sqrt[p]{T}$. 
In the presence of screening, the correlation length at infinite temperature
saturates and is equal to the screening length. Similarly, as seen by Eq.(\ref{gd}), the generalized Debye sreening length scales in precisely the same manner. 
In Eq.(\ref{xldebye}), we will comment on the relation between the two scales.

\secn{Generalizations}\label{gen}
Here we illustrate how our results can be generalized to systems which do not fall in the class of systems introduced in
Section \ref{sos}.

\subsecn{Disorder}
When Eq.(\ref{Ham}) is replaced by a system
with non-translationally invariant exchange couplings $V(\vec{x}, \vec{y})\equiv \langle \vec{x}|V|\vec{y}\rangle$,
then $V$ will be diagonal
in an orthonormal basis ($|\vec{u} \rangle$) different from the momentum space
eigenstates, i.e.,
$V|\vec{u}\rangle=v(\vec{u})|\vec{u}\rangle$.
Our derivation will be
identical in the $|\vec{u} \rangle$ basis.  In particular, Eq.(\ref{gkon'})
will be the same with $v(\vec{k})$ replaced by $v(\vec{u})$. 

\subsecn{Fluids}
Our results can be directly applied to fluids. In this case the spin at each site in Eq.(\ref{Ham})
may be replaced by the local mass density.
The pair structure factor $S(k)$ is the same as the Fourier space correlation function $G(k)$ \cite{marchtosi}.
For $r\neq0$, the pair distribution function $g(r)$ is related to the correlation function $G(r)$ defined above as
\begin{eqnarray}
g(r)=G(r)+1.
\end{eqnarray}
For $r=0$, $g(r)=0$.

\subsecn{General Multi-component Interactions}
In case of systems with multiple interacting degrees of freedom at each lattice site,
we have a similar result. We consider, for instance, the non-rotationally invariant $O(n)$ Hamiltonian,
\begin{eqnarray}
H = \frac{1}{2} \sum_{\vec{x}\neq\vec{y}}\sum_{a,b}V_{ab}(\vec{x},\vec{y})S_a(\vec{x})S_b(\vec{y}),
\label{Hamgen}
\end{eqnarray} 
where the interactions $V_{ab}(\vec{x}, \vec{y})$ depend on
the spin components $1 \le a,b \le n$ 
as well as the locations $\vec{x}$ and $\vec{y}$. By fiat,
in Eq.(\ref{Hamgen}), $V_{ab}(\vec{x}=\vec{y})=0$.
Non-rotationally symmetric interactions such as those of Eq. (\ref{Hamgen}) 
with a kernel $V_{ab}$ which is not proportional to the identity matrix in the internal spin space $1 \le a,b \le n$ 
appear in, e.g., Dzyaloshinsky-Moriya 
interactions \cite{DM}, 
isotropic \cite{kk} and non-isotropic compass \cite{zn2005}, Kugel-Khomskii
\cite{kk, budnik} and Kitaev type \cite{kitaev} models. Such interactions also appear 
in continuous and discretized non-abelian gauge backgrounds (and scalar products
associated with metrics of curved surfaces) used to describe metallic glasses and
cholesteric systems \cite{soccer, sadoc, nelson_book, sadoc1, sadoc2, nw, Nelson, sethna, sn, gilles}.
The lattice ``soccer ball'' spin model \cite{soccer} is precisely of the 
form of Eq. (\ref{Hamgen}).  Replicating the calculations leading to Eq.(\ref{g3}), 
for $\vec{x}\neq\vec{y}$, to ${\cal{O}}(1/T^2)$, we find that
\begin{eqnarray}
&&G_{ab}(\vec{x},\vec{y})=\langle S_a(\vec{x})S_b(\vec{y}) \rangle \nonumber\\
&&=-\frac{V_{ab}(\vec{x},\vec{y})}{k_BT}
+\frac{1}{(k_BT)^2}\sum_{c,\vec{z}}V_{ac}(\vec{x},\vec{z})V_{cb}(\vec{z},\vec{y}).
\label{gab}
\end{eqnarray}

\subsecn{Bose/Fermi gases}\label{bosefermi}
Here we discuss Bose/Fermi systems to illustrate the generality of our result from Eq.(\ref{gkon'}) 
We consider the Hamiltonian given by
\begin{eqnarray}
H=H_0+H_I,\nonumber
\end{eqnarray}
where
\begin{eqnarray}
H_0&=&\sum_{\vec{x}}\hat{\psi}^\dagger(\vec{x})\frac{{\bf p}^2}{2m}\hat{\psi}(\vec{x}),\nonumber\\
H_I&=&\frac{1}{2}\sum_{\vec{x},\vec{x}'}\hat{\rho}(\vec{x})V(\vec{x}-\vec{x}')\hat{\rho}(\vec{x}'),
\end{eqnarray}
with $\hat{\rho}(\vec{x})=\hat{\psi}^\dagger(\vec{x})\hat{\psi}(\vec{x})-\langle\hat{\psi}^\dagger\hat{\psi}\rangle_0$. 

Here and throughout, $\langle \cdot \rangle_0$ denotes an average with respect to $H_0$ (the ideal gas Hamiltonian).
The fields $\psi$ obey 
appropriate statistics(Bose-Einstein/Fermi-Dirac) depending on the system being studied.
The standard partition function is
\begin{eqnarray}
Z=Z_0\int D\eta(\vec{x},\tau)~e^{-\beta\Phi}.
\end{eqnarray}
Here, $\tau$ is the standard imaginary time coordinate ($ 0 \le \tau \le \beta$). 
$Z_0$ is the partition function of the non-interacting system described by $H_0$,
$\eta$-s are the dual fields after performing the HS transformation.
We can express $\Phi$ as 
\begin{eqnarray}
\Phi=-\frac{N^2}{2\beta^3}\int_0^\beta d\tau\sum_{\vec{x},\vec{x}'}\eta(\vec{x},\tau)V^{-1}(\vec{x}-\vec{x}')\eta(\vec{x}',\tau)\nonumber\\
-\frac{N}{\beta}\ln \left\langle T_\tau\exp\left(\frac{1}{\beta}\int_0^\beta d\tau\sum_{\vec{x}} \eta(\vec{x},\tau)\hat{\rho}(\vec{x},\tau)\right)\right\rangle_0.
\label{Phiht}
\end{eqnarray}
where $T_\tau$ is the (imaginary) time-ordering operator.
It is clear that
the factor of the partition function which controls high temperature behavior comes from the first term 
in $\Phi$. 
Thus, for small $\beta$ (high temperature), the distribution of the values of $\eta$ is sharply peaked around zero.
Also, for small $\beta$, the integrands of Eq.(\ref{Phi}) have little dependence on $\tau$. Therefore, at high temperature,
\begin{eqnarray}
\Phi=-\frac{N^2}{2\beta^2}\sum_{\vec{x},\vec{x}'}\eta(\vec{x})\left[V^{-1}(\vec{x}-\vec{x}')\right.\nonumber\\
\left.+\beta A(\vec{x}-\vec{x}')\right]\eta(\vec{x}'),
\label{Phi}
\end{eqnarray}
where $A(\vec{x}-\vec{x}')=\langle\rho(\vec{x})\rho(\vec{x}')\rangle_0=C\delta_{\vec{x},\vec{x}'}$, 
with $C=\rho_{0}^{2}$ being a constant. 
The correlation function for this system is defined as
$
G(\vec{x}-\vec{y})=\langle\rho(\vec{x})\rho(\vec{y})\rangle,
$
It is easy to show that written in terms of the dual variables,
\begin{eqnarray}
G(\vec{x}-\vec{y})=\left\langle\frac{f'(N\eta(\vec{x}))}{f(N\eta(\vec{x}))}
\frac{f'(N\eta(\vec{y}))}{f(N\eta(\vec{y}))}\right\rangle_d,
\end{eqnarray}
where
$
f(a)=\mbox{Tr}_{\rho(\vec{x})}~e^{a\rho(\vec{x})}
$
and, as before, 
$\langle\cdot\rangle_d$ denotes the average with respect to the dual fields $\eta$.
For small values of the $\eta$ variables (high temperature), we have in general,
$
G(\vec{x}-\vec{y})=C_0+C_1\langle\eta(\vec{x})\eta(\vec{y})\rangle_d,
$
with $C_0$ chosen such that $G(\vec{x})=C$ for $\vec{x}=0$
and $C_1$ defined by the statistics of $\rho$ and the form of the pair interaction $V$.
Therefore, we have, 
\begin{eqnarray}
G(\vec{k}) 
&=&C+\frac{C_1k_BT}{C[Cv(\vec{k})+k_BT]}\nonumber\\
& &\ \ \ \ -\frac{1}{N}\sum_{\vec{k}}\frac{C_1k_BT}{C[Cv(\vec{k})+k_BT]}.
\label{gkquant}
\end{eqnarray}
This is similar to the classical $O(n)$ correlation function in Fourier space [Eq.(\ref{gkon'})] 
We can easily generalize Eq.(\ref{gkquant}) for multi-component/polyatomic systems as in Eq.(\ref{gab}).
Applied to {\em scattering data} from such systems, our results may enable the determination
of effective unknown microscopic interactions that underlie the system. 
Similarly, replicating the same derivation, mutatis mutandis, for
quantum SU(2) spins $\vec{S} = (S_{x}, S_{y}, S_{z})$ in the coherent spin
representation leads to the high temperature result
of three-component  ($O(n=3)$) classical spins.
This illustrates the well known maxim that at high temperature, 
details may become irrelevant and systems ``become classical''. In a similar manner, at high
$T$, the details underlying the classical $O(n)$ model
(the $O(n)$ normalization constraints concerning a fixed value of $|\vec{S}(\vec{x})|$
for $n$ component vectors $\vec{S}(\vec{x})$ at all sites $\vec{x}$) effectively 
became irrelevant at high temperature -- the behavior for all 
$n$ was similar.

\secn{Approximate Methods} \label{approx}
The exact high temperature
results that we obtained for lattice spin systems 
and the generalizations that we discussed in 
Section \ref{gen} are, as we will show below, 
similar to those attained by several approximate methods.
This coincidence of our exact results with the more
standard and intuitive approximations enables
a better understanding from different approaches.
A corollary of what we discuss below is
that the divergence of the correlation lengths
in systems with long range interactions
in the high temperature limit (as in Section \ref{htcl})
appears in all of these standard approximations. 
However, as we illustrated earlier in our work,
and in Section \ref{htcl} in particular, this divergence
is not a consequence of a certain approximation but
is an exact feature of all of these systems in their high temperature
limit.

In what follows, we will specifically discuss 
 {\bf(i)} $\phi^{4}$ field theories, {\bf(ii)} the large $n$ limit, and 
{\bf{(iii)}}  the OZ approach for fluids
invoking the mean-spherical approximation (MSA) \cite{MSA}.

\subsecn{Ginzburg-Landau $\phi^4$-type theories} \label{gl}

In the canonical case, the free energy density of the $\phi^{4}$ theory is given by
\begin{eqnarray}
{\cal F}=\frac{1}{2}(\nabla\phi(\vec{x}))^2+\frac{1}{2}r\phi^2(\vec{x})+\frac{a}{4!}\phi^4(\vec{x}).
\label{ff}
\end{eqnarray}
A finite  value of $a$ corresponds to the ``soft-spin'' approximation
where the norm is in not constrained, 
$\langle \phi^2(\vec{x})\rangle\neq1$.
Here, $r=c(T-T_0)$, with $c$ a positive constant.
The partition function \cite{beta1} is
$Z=\int D\phi~e^{-F}$ where $F=\int{\cal F}~d^dx$
with $d$ the spatial dimension.
At high temperature, the correlator behaves in a standard way (the OZ form)
$\left\langle|\phi(\vec{k})|^2\right\rangle=\frac{1}{k^2+r}$.
The irrelevance of the $\phi^4$ term may, e.g., be seen by effectively setting 
$\phi^4(\vec{x})\to6\langle\phi^2(\vec{x})\rangle\phi^2(\vec{x})$
in the computation of the partition function.
As $\langle\phi^2(\vec{x})\rangle$ is small [in fact, from Fourier transforming the above, $\langle\phi^2(\vec{x})\rangle={\cal O}(1/T)]$, the $\phi^4$ term is smaller than the $(\nabla\phi)^2$ term in Eq.(\ref{ff}) by a factor of $a/T$
and therefore can be neglected.
When general two body interactions with an interaction kernel
$v(\vec{k})$ are present, we similarly have
$
\langle|\phi(\vec{k})|^2\rangle=\frac{1}{v(\vec{k})+r}.
$ Our result of Eq.(\ref{gkon'}) for interactions of arbitrary spatial range is new and illustrates
that suggestive results for the correlation lengths 
attained by soft spin approximations
are not far off the mark for general systems
in the high temperature limit. As far as we are aware,
the high temperature correlation length of general theories was not known to be 
similar to that suggested by various perturbative schemes
(including the $1/n$  \cite{ma_paper}
and $\epsilon$ expansions  \cite{wilson_fisher}). 

\subsecn{Correlation Functions in the large $n$ limit}\label{largeN}
We now provide a derivation of Eq.(\ref{gkon'}) 
as it applies in the large $n$ limit.
Long ago, Stanley \cite{stanley} demonstrated that the 
large $n$ limit of the $O(n)$ spins
is identical
to the spherical model first introduced by Berlin and Kac \cite{kac}.

The single component spherical model is given by the Hamiltonian,
\begin{eqnarray}
H = \frac{1}{2} \sum_{\vec{x}\neq\vec{y}}V(|\vec{x}-\vec{y}|)S(\vec{x})S(\vec{y}). 
\label{Ham_scal}
\end{eqnarray} 
The spins in Eq.(\ref{Ham_scal})
satisfy a single global (``spherical'') constraint,
\begin{eqnarray}
\sum_{\vec{x}} S^{2}(\vec{x}) = N,
\label{con}
\end{eqnarray}
enforced in its average value \cite{MSA} by a Lagrange multiplier $\mu$.
 This leads to the functional $H'=H+\mu N$ which renders the model quadratic (as both Eqs.(\ref{Ham_scal}, \ref{con}) are quadratic) and thus exactly solvable, see, e.g., (\cite{us}).

From the equipartition
theorem, for $T\ge T_c$, where no condensate is present, the Fourier space correlator 
\begin{equation}
G(\vec{k})=\frac{1}{N}\langle |s(\vec{k})|^2\rangle=\frac{k_BT}{v(\vec{k})+\mu}.
\label{sksq}
\end{equation}
The real space two point correlator is given by
\begin{eqnarray}
G(\vec{x}) \equiv \langle S(0) S( \vec{x}) \rangle = \frac{k_{B} T}{N}
\sum_{\vec{k}} \frac{e^{i\vec{k} \cdot \vec{x}}}
{v(\vec{k}) + \mu}.
\label{corr}
\end{eqnarray}
To complete the characterization of
the correlation functions at different temperatures, we note that the
Lagrange multiplier $\mu(T)$ is given by the implicit equation $1=
G(\vec{x}=0)$.
Thus,
\begin{equation}
\frac{k_{B} T}{N}
\sum_{\vec{k}} \frac{1}
{v(\vec{k}) + \mu}=1.
\label{g1mut}
\end{equation}
This implies that the temperature $T$ is a monotonic increasing function of $\mu$. 
Eq.(\ref{g1mut}) also implies that in the high temperature limit,
\begin{eqnarray}
\mu=k_BT.
\label{mukbt}
\end{eqnarray}
Taken together, Eqs.(\ref{sksq},\ref{mukbt}) yield Eq.(\ref{gkon'}) in the asymptotic high temperature limit.
For completeness, we briefly note what happens at low $T$ ($T<T_C$).
In the spherical model,
at  the critical temperature ($T_c$),
the Lagrange multiplier $\mu$ takes the value,
\begin{equation}
\mu_{min}=-\min_{\vec{k}}\{v(\vec{k})\}.
\label{mumin}
\end{equation}
For $T<T_c$, (at least) one mode $\vec{q}$ is macroscopically occupied,
the mode(s), $\vec{q}$ being occupied is one for which $v(\vec{k})$ is minimum.
The ``condensate fraction''  $\langle|s(\vec{q})|^2\rangle/N^2>0$.

\subsecn{Ornstein-Zernike Equation}\label{oze}
As noted earlier, application of
the MSA to the OZ equation
for fluids
reproduces similar results for the ``total correlation function'', $h(\vec{r})$.
This is defined as 
$h(\vec{r})=g(\vec{r})-1$, where $g(\vec{r})$ is the standard radial distribution function.
The OZ equation for a fluid with particle density $\rho$ is given by,
\begin{eqnarray}
h(\vec{r})=C(\vec{r})+\rho\int dr' C(\vec{r}-\vec{r'})h(\vec{r'}),
\end{eqnarray}
where $C(\vec{r})$ is the ``direct correlation function''. Using the MSA,
$C(\vec{r})=-\beta V(\vec{r})$
\cite{MSA}, we get in Fourier space,
\begin{eqnarray}
S(\vec{k})&=&\frac{k_BT}{\rho v(\vec{k})+k_BT}.
\end{eqnarray}
This is similar to our result for $G(\vec{k})$.
However, it is valid 
only for systems in which the MSA is a good approximation.

\secn{Conclusions} \label{conc}
{\bf (i)} We derived a universal form 
for high temperature
correlators in general $O(n)$
theories.  This enables the {\em extraction of unknown microscopic interactions}
from measurements of high temperature correlation function.

{\bf(ii)} We discovered {\em divergent} correlation lengths in systems
with long range interactions in the high temperature limit. 
This divergence is replaced by a saturation 
when the long range interactions
are screened.

{\bf(iii)} We introduced {\em generalized Debye lengths} associated with such divergent correlation lengths.


\begin{appendix}

\secn{High temperature series expansion of the correlation function}\label{htseries}
We now outline in detail how we may obtain a high temperature ($T$)
series expansion of the correlation function to arbitrary order
for a general system with translational invariance.
The result provided 
above
was derived to order ${\cal{O}}(1/T^{2})$.
This and the results we present below are valid in the high temperature phase of general lattice (spin or other) 
and continuum systems.
However, it will hold in lower temperature phases of the system provided we can analytically continue
to those phases from the high temperature phase, i.e. attain those phases without having a phase transition.
It is also worth mentioning that since we set the temperature to be arbitrarily high, the density does not have 
to be small as is assumed in methods derived from Mayer's cluster expansion for fluids.
Our result is therefore valid for the high temperature phase of any system.
In general, the long range character of the interactions
will not enable us to invoke many of the simplifying elegant tricks present elsewhere.
For instance, the counting of connected contours and loops \cite{loop, Binney}
that appear in high temperature series expansion involving nearest 
neighbor interactions cannot be applied here. 

We can perform the high temperature series expansion directly in the original spin space.
However, we find it easier to make a transformation to a dual space where our Boltzmann weights become 
Gaussian in the high temperature limit. 

The correlation function of the original theory can be expressed in terms of the
correlation function (and higher moments) of the dual theory
-- we employ that in our calculation.
The dual theory to a nearest neighbor ferromagnetic system
is a Coulomb gas.
Nearest neighbor ferromagnetic system in dimensions $d>2$ at low $T$ has an ordered phase
and a small correlation length (correlation length diverges at $T=T_{c}$).
This does {\em not} imply that the Coulomb system has a small correlation
length at high temperature. $O(n)$ constraints become faint at high
$T$ in the dual theory whereas in the exact Coulomb gas at
high T, the $O(n)$ constraints are there. The same also
applies for a soft spin realization of the Coulomb gas
where $\exp[-\beta u(S^{2}-1)^{2}]$ which is zero
as $\beta \to \infty$ (or $T \to 0$) unless
$S^{2}=1$ everywhere. By contrast in the exact
dual theory at high temperature, the relative strength
of the $O(n)$ constraints becomes negligible relative compared to
the ``interaction'' term containing $(\beta V)^{-1}$.
Even though we can  ignore $\beta$ prefactors
when $\beta = {\cal{O}}(1)$ and consider dual theories
and soft spin realization we cannot ignore
the $T$ dependence at high T about
the infinite $T$ disordered limit.
Otherwise we get a contradiction as our
exact calculation with the exact dual 
theory (containing the $T$ dependent
prefactors) shows.

We will keep things general and
perform the simple series expansion of the dual Hamiltonian $H_d$ in Eq.(\ref{dual}). 
\begin{eqnarray}
H_{d}&=&-\frac{N^2}{2\beta^2}\sum_{\vec{x}, \vec{y}}V^{-1}(\vec{x}-\vec{y})\vec{\eta}(\vec{x})\cdot\vec{\eta}(\vec{y})\nonumber\\
& &-\frac{1}{\beta}\sum_{\vec{x}}\ln\left(\frac{I_{n/2-1}(\sqrt{n}N|\vec{\eta}(\vec{x})|)}{(\sqrt{n}N|\vec{\eta}(\vec{x})|)^{n/2-1}}\right),\nonumber\\
&=&-\frac{N^2}{2\beta^2}\sum_{\vec{x}, \vec{y}}V^{-1}(\vec{x}-\vec{y})\vec{\eta}(\vec{x})\cdot\vec{\eta}(\vec{y})\nonumber\\
& &-\frac{N^2}{2\beta}\sum_{\vec{x}}\vec{\eta}(\vec{x})\cdot\vec{\eta}(\vec{x})\nonumber\\
& &+\frac{N^4}{4(n+2)\beta}\sum_{\vec{x}}\left[\vec{\eta}(\vec{x})\cdot\vec{\eta}(\vec{x})\right]^2+...
\label{dual_exp}
\end{eqnarray}
In Eq.(\ref{dual_exp}), the interaction $V$ should be thought of as a translationally invariant matrix.
That is, in a Dirac type notation, 
$\langle \vec{x} | V| \vec{y} \rangle = V(\vec{x}- \vec{y})$. In Eq.(\ref{dual_exp}), 
$V^{-1}$ is the inverse Fourier transform of $1/v(\vec{k})$, where $v(\vec{k})$ is the Fourier transform of $V(\vec{x})$
 \cite{constant}.

 Next, we separate $H_{d}$ 
into a quadratic part $H_{d0}$ and higher order (interaction type) terms which we denote by $\Delta H$.
That is,
\begin{eqnarray}
H_{d0}&=&-\frac{N^2}{2\beta^2}\sum_{\vec{x}, \vec{y}}V^{-1}(\vec{x}-\vec{y})\vec{\eta}(\vec{x})\cdot\vec{\eta}(\vec{y})\nonumber\\
& &-\frac{N^2}{2\beta}\sum_{\vec{x}}\vec{\eta}(\vec{x})\cdot\vec{\eta}(\vec{x})\\
\Delta H&=&\frac{N^4}{4(n+2)\beta}\sum_{\vec{x}}\left[\vec{\eta}(\vec{x})\cdot\vec{\eta}(\vec{x})\right]^2+...~.
\end{eqnarray}

The expectation value of any quantity $X$ may be computed by
\begin{eqnarray}
\langle X\rangle_d&=&\frac{\langle Xe^{-\beta\Delta H}\rangle_{d0}}{\langle e^{-\beta\Delta H}\rangle_{d0}},\\
&=&\langle X \rangle_{d0} - \beta \left[\langle X \Delta H \rangle_{d0}- \langle X  \rangle_{d0} \langle \Delta H \rangle_{d0}  \right] \nonumber\\
& & + \frac{\beta^2}{2!} \left[\langle X (\Delta H)^2  \rangle_{d0} -2\langle X\Delta H \rangle_{d0}\langle \Delta H \rangle_{d0} \right.\nonumber\\
& & \left.+2 \langle X  \rangle_{d0} \langle \Delta H \rangle_{d0}^2 - \langle X  \rangle_{d0} \langle (\Delta H)^2 \rangle_{d0} \right] \nonumber\\
& & +...~, 
\end{eqnarray}
where $\langle\cdot\rangle_{d0}$ represents the expectation value calculated with 
the Boltzmann weight associated with the Hamiltonian $H_{d0}$.
We may retain terms
to arbitrary order in $\eta^2$ (or corresponding order in $1/T$).
Eq.(\ref{gtan}) 
can be expanded to arbitrary order in $\eta^2$ where we
rewrite all expectation values with respect to the Hamiltonian $H_{d0}$.
The terms become expectation values of a product of an even number of $\eta$ fields with
respect to the quadratic Hamiltonian $H_{d0}$.
We can then use the Wick's theorem to compute the 
expectations with respect to $H_{d0}$ to all orders.
To order $1/T^3$ we obtain for $\vec{x}\neq0$,
\begin{eqnarray}
\label{g3}
&&G(\vec{x})=-\frac{V(\vec{x})}{k_BT}+\nonumber\\
&&\frac{1}{(k_BT)^2}\left[\sum_{\vec{z}}V(\vec{z})V(\vec{x}-\vec{z})-2V(0)V(\vec{x})\right]\nonumber\\
+&&\frac{1}{(k_BT)^3}\left[-\sum_{\vec{y},\vec{z}}V(\vec{y})V(\vec{z})V(\vec{x}-\vec{y}-\vec{z})+\right.\nonumber\\
&&2V(\vec{x})\sum_{\vec{z}}V(\vec{z})V(-\vec{z})+3V(0)\sum_{\vec{z}}V(\vec{z})V(\vec{x}-\vec{z})\nonumber\\
&&\left.-5(V(0))^2V(\vec{x})-2\frac{(V(\vec{x}))^3}{n+2}\right].
\end{eqnarray}
As a brief aside, we note that from the fluctuation dissipation theorem the susceptibility
$
\chi=\beta\sum_{\vec{x}}G(\vec{x}).
$
At asymptotically high temperature, $G(\vec{x}) \simeq  \delta_{\vec{x},0}$
giving rise to Curie's law, $\chi\propto1/T$.
The terms in Eq.(\ref{g3}) lead to higher order corrections.
To next order, 
$
\chi=\frac{1}{k_B(T-\theta_C)}
$
with the Curie temperature 
$
\theta_C=\sum_{\vec{x}\neq0}V(\vec{x})
$
in the weak coupling limit.
Thus far, in the literature, the Curie-Weiss form was invoked to ascertain whether a given
system has dominantly ferromagnetic or anti-ferromagnetic interactions (sign of $\theta_C$)
and their strength ($|\theta_C|$).
We see that by not focusing solely on $\chi=\beta G(\vec{k}=0)$ but rather on the scattering function $G(\vec{k})$
for all $\vec{k}$, we can in principle deduce the interaction $v(\vec{k})$ and hence $V(\vec{x})$.

In Fourier space, the real space convolutions become momentum space products and vice versa.
Eq.(\ref{g3}) then reads
\begin{eqnarray}
&&G(\vec{k})=1-\frac{v(\vec{k})}{k_BT}+\frac{1}{(k_BT)^2}\left[(v(\vec{k}))^2-2V(\vec{x}=0)v(\vec{k})\right]\nonumber\\
&&+\frac{1}{(k_BT)^3}\left[-(v(\vec{k}))^3+\frac{2v(\vec{k})}{N}\sum_{\vec{k_1}}(v(\vec{k_1}))^2\right.\nonumber\\
&&+3V(\vec{x}=0)(v(\vec{k}))^2-5(V(\vec{x}=0))^2v(\vec{k})-\frac{2}{N^2(n+2)}\times\nonumber\\
&&\left.\sum_{\vec{k_1},\vec{k_2}}v(\vec{k_1})v(\vec{k_2})v(\vec{k}-\vec{k_1}-\vec{k_2})\right]-G_1(0),
\label{g2t}
\end{eqnarray}
where $G_1(0)$ is the value obtained by inserting $\vec{x}=0$ in Eq.(\ref{g3}).
It should be noted that the real space correlation function 
cannot change if we shift the on-site interaction
$V(\vec{x}=0)$ which is equivalent to a uniform shift to $v(\vec{k})$ for all $k$.
This is because
the ${\cal{O}}(n)$ spin is normalized --  $|\vec{S}(\vec{x})|^2=n$ at all sites $\vec{x}$.
This invariance
to a constant shift holds for all $T$
and consequently to any order in $1/T$, the coefficients
must be invariant to a global shift in $v(k)$. Amongst other
things, we earlier invoked this invariance \cite{explain_shift} 
to shift $v(\vec{k})$ to enable a  HS
transformation in the cases for which initially $v(\vec{k})>0$ for some values 
of $\vec{k}$. We can, of course, invoke this invariance also here 
to obtain the above high temperature series expansion with a well 
defined HS dual. The final results, as we re-iterated above
are invariant under this shift as is also manifest in our series expansion
in powers of $1/T$. Although obvious, we note that the expansion in Eq.(\ref{g2t})
is performed in power of $1/T$ involving $v(k)$ for real vectors $\vec{k}$.
In examining the correlation lengths via contour integration in the complex
$k$ plane, the corresponding $v(k)$ may be extended for
complex $k$. 

We see from the expansion in Eq.(\ref{g2t}) that already
to ${\cal{O}}(1/T)$, it is also clear that the lengthscales of the system (which are
determined by the poles of the Fourier space correlation function) are governed by the
poles of $v(\vec{k})$ in the complex $\vec{k}$ space.
Thus, if, e.g., $v(\vec{k})=1/(k^2+\lambda^{-2})$, the correlation length tends to $\lambda$ at high temperature.
It therefore must diverge for a system with no screening.

In cases where the correlation function is known from some experimental technique or otherwise, the series expansion for 
the correlation function can be inverted to arbitrary order to obtain the pairwise interactions. To ${\cal{O}}(1/T^2)$,
for non-zero separation $\vec{x}$, the potential function is given by,
\begin{eqnarray}
\label{last}
V(\vec{x})=-k_BT\left[G(\vec{x})-{\sum_{\vec{z}}}'G(\vec{z})G(\vec{x}-\vec{z})\right.\nonumber\\
+{\sum_{\vec{y},\vec{z}}}'G(\vec{y})G(\vec{z})G(\vec{x}-\vec{y}-\vec{z})\nonumber\\
\left.+2G(\vec{x}){\sum_{\vec{z}}}'G(\vec{z})G(-\vec{z})-\frac{2\left(G(\vec{x})\right)^3}{n+2}\right].
\end{eqnarray}
The prime indicates that the sum excludes terms containing $G(0)$.
As is evident from our earlier results and discussion, in Eq.(\ref{last}), 
each correlation function $G(\vec{x})$ is of order
$(1/T)$. 

We re-iternate that as in our discussion in Section \ref{approx},  our results
for lattice $O(n)$ spin models
match with the leading order behavior at high temperature 
obtained from several standard approximate theories
based on Mayer's cluster expansion derived for liquid systems, e.g., Born-Green theory \cite{BG} 
and OZ theory
with Percus-Yevick approximation \cite{PY} or MSA \cite{MSA}.
As implicit above, 
our $1/T$ expansion
can indeed
be extended to systems in which the liquid and the gas phase are not separated by a
phase transition,
e.g.,
for pressures larger than the pressure at the liquid-vapor critical point.
As further noted in Section \ref{approx}, various approximations
also suggest that at high temperature, the correlation
length may match the length-scale characterizing of the interaction potential
and, in particular, would diverge in systems having long range interactions
(as we have established).
\\


\secn{Relation between the generalized Debye lengths and divergence of the high temperature correlation lengths}
\label{debye_rel}
An intuitive approximate approach for the understanding of the rigorous yet seemingly paradoxical result
that we report in this work- that of the divergence of the correlation lengths in the high
temperature limit of systems with long range interactions is afforded by the OZ framework.
Specifically, in the language of OZ approximations, the ``total'' high temperature correlation function 
is the same as the ``direct'' correlation function (see, e.g., \cite{marchtosi} (section 2.6)
for the definition of the ``direct'' OZ correlation functions) 
and behaves as
\begin{eqnarray}
G(\vec{x})\sim-\beta V(\vec{x})
\end{eqnarray}
for $\vec{x}\neq0$.
Thus, if the potential is screened beyond a distance $\lambda$, the correlation length
approaches $\lambda$ at high temperature.
That is, if we have an effective interaction
resulting, e.g., from higher order effects in $1/T$, 
such as that leading to the Debye screening length ($\lambda_{D}$) in Coulomb systems 
(and generalizations introduced earlier in Eq.(\ref{gd})),
then at high temperature, the correlation length 
\begin{eqnarray}
\label{xldebye}
\xi \xrightarrow[T\to\infty]{} \lambda_D.
\end{eqnarray}
This is a particular case of Eq.(\ref{xilam}). 

To ${\cal{O}}(1/T^2)$, 
Eq.(\ref{g2t}) is identical to Eq.(\ref{gkon'}). 
The poles of $G$ in the complex $k$ plane can,
of course, be computed to by finding those of Eq.(\ref{gkon'}) 
or considering those directly of Eq.(\ref{g2t}): both give rise to the same answer
as they must.

\end{appendix}

\bibliography{mybiblio}

\begin{thebibliography}{46}
\expandafter\ifx\csname natexlab\endcsname\relax\def\natexlab#1{#1}\fi
\expandafter\ifx\csname bibnamefont\endcsname\relax
  \def\bibnamefont#1{#1}\fi
\expandafter\ifx\csname bibfnamefont\endcsname\relax
  \def\bibfnamefont#1{#1}\fi
\expandafter\ifx\csname citenamefont\endcsname\relax
  \def\citenamefont#1{#1}\fi
\expandafter\ifx\csname url\endcsname\relax
  \def\url#1{\texttt{#1}}\fi
\expandafter\ifx\csname urlprefix\endcsname\relax\def\urlprefix{URL }\fi
\providecommand{\bibinfo}[2]{#2}
\providecommand{\eprint}[2][]{\url{#2}}

\bibitem[{ma_({\natexlab{a}})}]{ma_book}
\bibinfo{note}{S. K. Ma, {\em Modern Theory of Critical Phenomena},
  Addison-Wesley (1976)}.

\bibitem[{wil({\natexlab{a}})}]{wilson}
\bibinfo{note}{K. G. Wilson, Phys. Rev. B {\bf 4}, 3174 (1971)}.

\bibitem[{fis()}]{fisher}
\bibinfo{note}{M. E. Fisher, Rep. Prog. Phys. {\bf 30}, 615 (1967)}.

\bibitem[{deb()}]{debye_huckel}
\bibinfo{note}{P. Debye and E. Huckel, Physik Z, {\bf 24}, 185 (1923)}.

\bibitem[{sta({\natexlab{a}})}]{stanleykaplan}
\bibinfo{note}{G. S. Rushbrooke and P. J. Wood, Mol. Phys, {\bf 1}, 257 (1958);
  H. E. Stanley and T. A. Kaplan, Phys. Rev. Lett, {\bf 16}, 981 (1966)}.

\bibitem[{hua()}]{huang}
\bibinfo{note}{K. Huang, {\em Statistical Mechanics}, John Wiley and sons
  (1963,1987)}.

\bibitem[{sta({\natexlab{b}})}]{stanley}
\bibinfo{note}{H. E. Stanley, Phys. Rev. {\bf 176}, 2, 718 (1968).}

\bibitem[{kob()}]{kob}
\bibinfo{note}{J. Horbach, W.Kob, and K. Binder, Phil. Mag. B {\bf 77}, 2, 297
  (1998)}.

\bibitem[{SRL()}]{SRL}
\bibinfo{note}{R. L. Stratonovich, Soviet Physics Doklady {\bf 2}, 416 (1958);
  J. Hubbard, Phys. Rev. Lett. {\bf 3}, 77 (1959)}.

\bibitem[{iva()}]{ivanchenko}
\bibinfo{note}{Y. M. Ivanchenko and A. A. Lisyansky, {\em Physics of Critical
  Fluctuations}, Springer-Verlag (New York) (1995)}.

\bibitem[{exp()}]{explain_shift}
\bibinfo{note}{The integrals in Eq.(\ref{part1}) converge if $v(\vec{k})$ is
  negative for all real $\vec{k}$. This can be achieved for all systems with
  finite range or screened interactions (with an arbitrarily large screening
  length) when $v(\vec{k})$ is bounded from above. As
  $(\vec{S}(\vec{x}))^{2}=n$ for all $\vec{x}$, the kernel $v(\vec{k})$ for all
  $\vec{k}$ can be trivially made negative before performing the HS
  transformation by $V(\vec{x})\to V(\vec{x})+a\delta_{\vec{x},0}$ which merely
  shifts the energy by an additive constant, $H\to H+anN/2$, without changing
  any averages.}

\bibitem[{\citenamefont{Henderson}(1974)}]{henderson}
\bibinfo{author}{\bibfnamefont{R.~L.} \bibnamefont{Henderson}},
  \bibinfo{journal}{Physics Letters A} \textbf{\bibinfo{volume}{49}},
  \bibinfo{pages}{197 } (\bibinfo{year}{1974}).

\bibitem[{wha()}]{what_is_long?}
\bibinfo{note}{By ``long range'' we refer to systems for which $V(\vec{x})$ is
  not exactly zero for arbitrary large $\vec{x}$. In $d$ spatial dimensions,
  for large distances, $|V(\vec{x})| \propto \exp(-|\vec{x}|/\lambda)/x^{a}$
  with $0\le a\le d$ and $\lambda$ is a ``screening length'' which can in
  principle be arbitrarily large.}

\bibitem[{neu()}]{neutral}
\bibinfo{note}{By ``charge neutrality'', we mean that $\sum_{\vec{y}}
  \vec{S}(\vec{y}) =0$ and, consequently, $\langle \vec{S}(\vec{x}) \cdot
  \sum_{\vec{y}} \vec{S}(\vec{y}) \rangle
  =0\implies\sum_{\vec{z}}G(\vec{z})=0$.}

\bibitem[{sau()}]{saurish}
\bibinfo{note}{S. Chakrabarty and Z. Nussinov, arXiv:0906.5381}.

\bibitem[{us()}]{us}
\bibinfo{note}{L. Chayes et al. 
  G. Tarjus, Physica A {\bf 225}, 129 (1996).}

\bibitem[{zoh()}]{zohar}
\bibinfo{note}{Z. Nussinov , J. Rudnick, S. A. Kivelson, and L. N. Chayes,
  Phys. Rev. Letters 83, 472 (1999).}

\bibitem[{dip({\natexlab{a}})}]{dipole1}
\bibinfo{note}{A. Giuliani, J. L. Lebowitz and E. H. Lieb, Phys. Rev. B {\bf
  76}, 184426 (2007); Phys. Rev. B {\bf 74}, 064420 (2006).}

\bibitem[{dip({\natexlab{b}})}]{dipole2}
\bibinfo{note}{A. Vindigni, N. Saratz, O. Portmann, D. Pescia, and P. Politi,
  Phys. Rev. B {\bf 77}, 092414 (2008)}.

\bibitem[{dip({\natexlab{c}})}]{dipole3}
\bibinfo{note}{C. Ortix, J. Lorenzana, and C. Di Castro, Phys. Rev. B {\bf 73},
  245117 (2006)}.

\bibitem[{mar()}]{marchtosi}
\bibinfo{note}{N. H. March and M. P. Tosi, {\em Atomic Dynamics in Liquids},
  Dover Publications (1976)}.

\bibitem[{DM()}]{DM}
\bibinfo{note}{I. Dzyaloshinsky, J. Phys. Chem. Solids {\bf 4}, 241 (1958); T.
  Moriya, Phys. Rev. {\bf 120}, 1, 91 (1960)}.

\bibitem[{kk()}]{kk}
\bibinfo{note}{K. I. Kugel and D. I. Khomskii, Sov. Phys. Usp. {\bf 25} 231
  (1982)}.

\bibitem[{zn2()}]{zn2005}
\bibinfo{note}{Z. Nussinov and E. Fradkin, Phys. Rev. B {\bf 71}, 195120
  (2005)}.

\bibitem[{bud()}]{budnik}
\bibinfo{note}{R. Budnik and A. Auerbach, Phys. Rev. Lett. {\bf 93}, 187205,
  (2004)}.

\bibitem[{kit()}]{kitaev}
\bibinfo{note}{A. Kitaev, Annals of Physics {\bf 321}, 2 (2006)}.

\bibitem[{soc()}]{soccer}
\bibinfo{note}{Z. Nussinov, Phys. Rev. B {\bf 69} 014208 (2004)}.

\bibitem[{sad({\natexlab{a}})}]{sadoc}
\bibinfo{note}{R. Mosseri and J. F. Sadoc, J. Phys. Lett. (Paris) {\bf 45},
  L827 (1984); F. Sadoc and F. Mosseri {\em Amorphous Materials} edited by V.
  Vitek (American Institute of Metallurgical Engineering, New York, 1983), p.
  111 ; J. F. Sadoc and R. Mosseri, {\em Geometrical Frustration}, Cambridge
  University Press (1999), and references therein}.

\bibitem[{nel()}]{nelson_book}
\bibinfo{note}{D. R. Nelson, {\em Defects and Geometry in Condensed Matter
  Physics}, Cambridge University Press (2002)}.

\bibitem[{sad({\natexlab{b}})}]{sadoc1}
\bibinfo{note}{J. F. Sadoc, J. Phys. Lett. {\bf 44}, L707 (1983)}.

\bibitem[{sad({\natexlab{c}})}]{sadoc2}
\bibinfo{note}{J. F. Sadoc and J. Charvolin, J. Phys. {\bf 47}, 683 (1986)}.

\bibitem[{nw()}]{nw}
\bibinfo{note}{D. R. Nelson and M. Widom, Nucl. Phys. B {\bf 240}, 113 (1984)}.

\bibitem[{Nel()}]{Nelson}
\bibinfo{note}{D.R. Nelson, Phys. Rev. B {\bf 28}, 5515 (1983)}.

\bibitem[{set()}]{sethna}
\bibinfo{note}{J.P. Sethna, Phys. Rev. B {\bf 31}, 6278 (1985)}.

\bibitem[{sn()}]{sn}
\bibinfo{note}{S. Sachdev and D.R. Nelson, Phys. Rev. B {\bf 32}, 1480 (1985)}.

\bibitem[{gil()}]{gilles}
\bibinfo{note}{G. Tarjus, S. A. Kivelson, Z. Nussinov and P. Viot, J. Phys
  Cond. Matt. {\bf 17}, 50 (2005).}

\bibitem[{MSA()}]{MSA}
\bibinfo{note}{J. L. Lebowitz and J. K. Percus, Phys. Rev. {\bf 144}, 1
  (1966)}.

\bibitem[{bet()}]{beta1}
\bibinfo{note}{Here, we adhere to the prevalent practice of effectively setting
  $\beta=1$ (with the temperature dependence relegated solely to the prefactor
  $r$ of Eq.(\ref{ff})).}

\bibitem[{ma_({\natexlab{b}})}]{ma_paper}
\bibinfo{note}{S. K. Ma, Phys. Rev. A {\bf 7}, 2172 (1973)}.

\bibitem[{wil({\natexlab{b}})}]{wilson_fisher}
\bibinfo{note}{K. G. Wilson and M. E. Fisher, Phys. Rev. Lett. {\bf 28}, 240
  (1972)}.

\bibitem[{kac()}]{kac}
\bibinfo{note}{T. H. Berlin and M. Kac, Phys. Rev. {\bf 86}, 821 (1952)}.

\bibitem[{loo()}]{loop}
\bibinfo{note}{By ``connected contours'', we refer to contours whose links
  connect nearest neighbor sites. For general interactions, the individual
  links may have arbitrary length and weight. Albeit some notable differences,
  this is, essentially, reflected in our expansion where $V(\vec{x}-\vec{y})$
  is, in general, a link connecting sites $\vec{x}$ and $\vec{y}$.}

\bibitem[{Bin()}]{Binney}
\bibinfo{note}{For an introduction to the more elegant standard methods for
  systems with short range interactions, see any one of the many excellent
  textbooks on the subject, e.g., J. J. Binney, A. J. Fisher, and M. Newman,
  {\em The Theory of Critical Phenomena}, Oxford University Press (1992).}

\bibitem[{con()}]{constant}
\bibinfo{note}{The general Hamiltonian that $V$ encodes allows for (constant)
  on-site energies is a trivial generalization of Eq.(1): 
  $H=\frac{1}{2} \sum_{\vec{x},\vec{y}} V(|\vec{x}-\vec{y}|)\vec{S}(\vec{x})
  \cdot \vec{S}(\vec{y})$ that also includes a sum over sites $\vec{x} =
  \vec{y}$ (i.e., allows for $V(\vec{x}=0)\neq 0$).}

\bibitem[{BG()}]{BG}
\bibinfo{note}{M. Born and H. S. Green, Proc. Roy. Soc. London A {\bf 188}, 10
  (1946)}.

\bibitem[{PY()}]{PY}
\bibinfo{note}{J. K. Percus and G. J. Yevick, Phys. Rev. {\bf 110}, 1 (1958)}.

\end{thebibliography}

\end{document}